# Magneto-optical Control of Ordering Kinetics and Vacancy Behavior in Fe-Al Thin Films Quenched by Laser

*I. Yu. Pashenkin*[1], *D. A. Tatarskiy*[1,2], *S. A. Churin*[1], *A. N. Nechay*[1], *M. N. Drozdov*[1], *M. V. Sapozhnikov*[1,2], *N. I. Polushkin*[1] ∗

[1]Institute for Physics of Microstructures, Russian Academy of Sciences, Akademicheskaya St. 7, 603087, Nizhny Novgorod, Russia

[2] Faculty of Physics, Lobachevsky State University, Nizhny Novgorod 603950, Russia

∗Corresponding author: nip@ipmras.ru

Keywords: order-disorder phase transitions, crystal defects, disordered alloys, magnetic thin films, laser ablation, magneto-optical Kerr effect

## Abstract

One of issues arising in materials science is a behavior of non-equilibrium point defects in the atomic lattice, which defines the rates of chemical reactions and relaxation processes as well as affects the physical properties of solids. It is previously theoretically predicted that melting and rapid solidification of metals and alloys provides a vacancy concentration in the quenched material, which can be comparable to that quantity at the point of melting. Here, the vacancy behavior is studied exerimentally in thin films of the near equiatomic Fe-Al alloy subjected to nanosecond laser annealing with intensities up to film ablation. The effects of laser irradiation are studied by monitoring magneto-optically the ordering kinetics in the alloy at the very ablation edge, within a narrow (micron-scale) ring-shaped region around the ablation zone. Quantitatively, the vacancy supersaturation in the quenched alloy has been estimated by fitting a simulated temporal evolution of the long-range chemical order to the obtained experimental data. Laser quenching (LQ) of alloys and single-element materials would be a tool for obtaining novel phase states within a small volume of the crystal.

**Graphical Abstract**

One of challenges in materials physics and technology is to accelerate chemical ordering in substitutional alloys occurring through vacancies in the atomic lattice to improve their functional properties. This work addresses an attempt to solve this issue by generating extra vacancies in alloys using nanosecond laser irradiation. Methodologically, we explore the ordering kinetics in near equiatomic Fe-Al alloys using the magneto-optical Kerr effect in combination with structural studies.

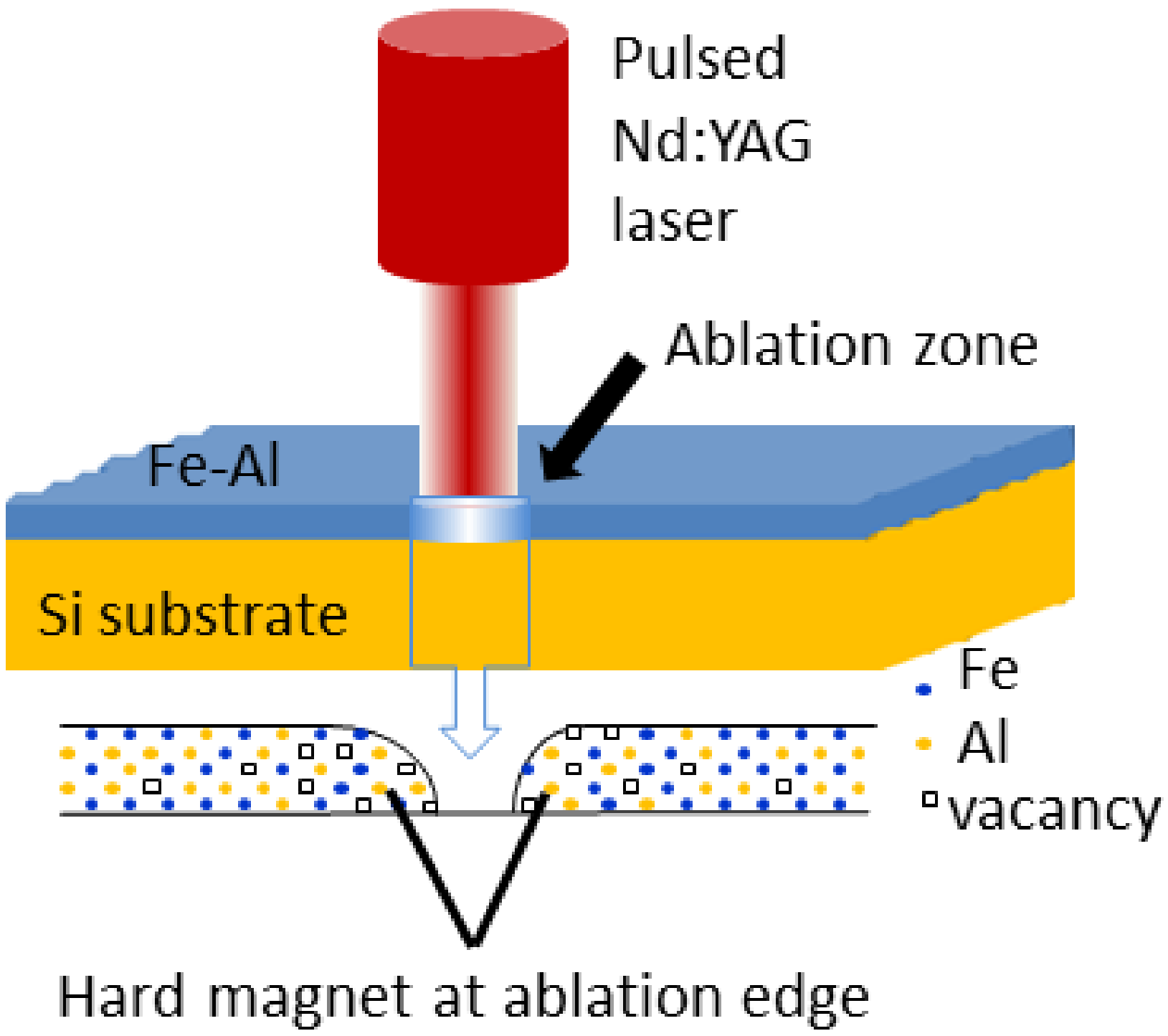

## 1. Introduction

Heating and melting of a material with a short laser pulse followed by rapid cooling, known as laser quenching (LQ), [1-3] is a proven method to effectively freeze it at a non-equilibrium state, which can be used in various implications [4-9]. In particular, it is of interest to produce a crystalline system with high densities of non-equilibrium structural defects such as vacancies and to explore how the defects behave, affecting defect-mediated atomic transport [10-17], phase transformations [18] and physical properties of functional alloys [19-22], compounds [23-26], and single-element materials [27, 28]. Theoretically, it was shown [29, 30] that, under conditions of strong undercooling of a material melted by a short laser pulse, the fast movement of the solid-liquid interface and lowered vacancy mobility prevent the equilibration of vacancy concentration $c_v$ in the solidified region, thus creating a vacancy supersaturation. If the cooling of the solidified material is rapid enough [8], excess vacancies do not have time to migrate and annihilate at their sinks. For the simulations [29, 30], the systems with an equilibrium vacancy concentration of $c_{\mathrm{eq}}$~$10^{-3}$ at the melting point $T_m$ where chosen, which dictated the level of excess vacancy concentration during cooling of the material. As for experiments, detection and characterization of non-equilibrium vacancies produced by laser were rare for years. An obstacle for developing this research field with employing short-pulse laser beams is that a laser spot is too small and/or a laser beam is too non-uniform over its cross-section to evaluate the quantity of $c_v$ and its temporal evolution using established methods such as positron annihilation [31], differential dilatometry [32], and residual electrical resistance [33].

The study presented here addresses to the effect of nanosecond laser irradiation on the ordering kinetics in near equiatomic Fe-Al alloys (~55 at. % Fe). It is well established [7-9, 34-48] that these alloyed systems exhibit transformations between chemically disordered A2 and ordered B2 states, which occur through *vacancies* in the atomic lattice [38-42], as illustrated in Figure 1. The critical temperature for order-disorder transformations in the equiatomic Fe-Al bulk is $T_c$=1590 K, which is very close to $T_m$ [36]. The binary Fe-Al alloy was chosen as having a relatively high level of $c_{\mathrm{eq}}(T)$, which reaches ~0.1 at $T_m$ [41]. In the context of these studies, it is also important that this alloy exhibits transitions between the paramagnetic and ferromagnetic states, [7-9, 34, 35, 42-48] which correspond to the B2 and A2 atomic structures, respectively, as also illustrated in Figure 1. Using magneto-optical Kerr effect (MOKE) magnetometry, with the light beam properly focused to probe small regions on the surface of the sample, we have found that, in the films quenched at laser intensities sufficient for their partial ablation, there is (1) enhancement of the magnetic response, [49] (2) the kinetics of ordering accelerates under aging at moderate temperatures *T* [50] and (3) there is an increase in magnetic coercivity that persists in the aged samples. Comparison of the obtained temporal (*t*) dependences of long-range chemical order $0<S<1$ with simulated ones [40, 51] allows for estimating the concentration of laser-generated vacancies.

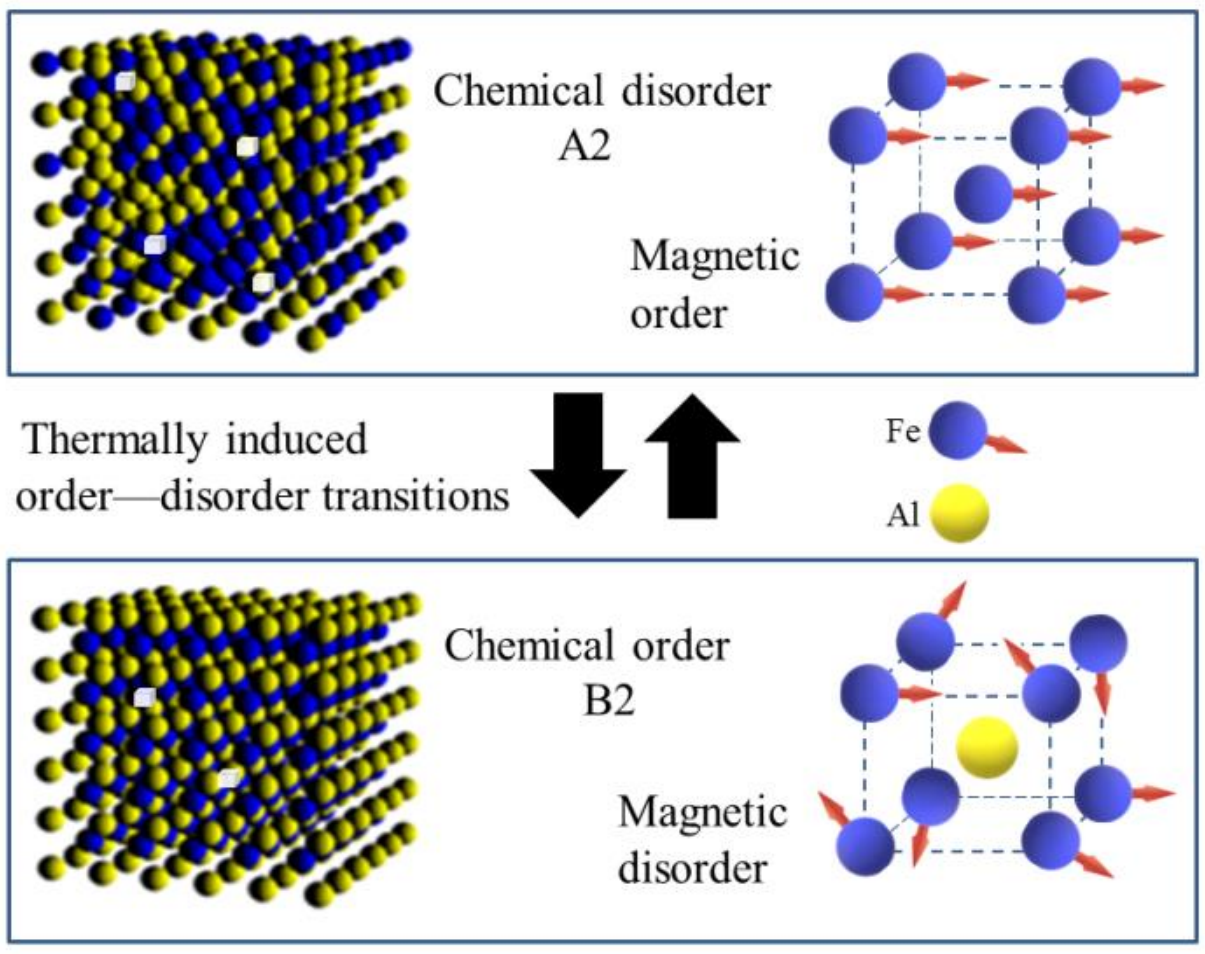

**Figure 1:** Feasible phase states in the body-centered cubic equiatomic Fe-Al structure, where chemical order-disorder phase transformations occur through vacancies in the atomic lattice (white boxes): (top) A metastable chemically disordered A2 but magnetically ordered (ferromagnetic) state and (bottom) the ground chemically ordered B2 but magnetically disordered (paramagnetic) state. In the A2 state, the center of the unit cell can be occupied by an Fe atom (antisite defect), and so, the alloy becomes ferromagnetically ordered, with the magnetic moments (represented by red arrows) aligned in the same direction. The average concentration of Fe atoms in the alloy is sufficient for magnetic percolation to occur.

## 2. Results and Discussion

### 2.1. Structural characterization of the as-grown and thermally annealed samples

Using transmission electron microscopy (TEM), it is found that thin-film (50 nm) Fe-Al samples grown at room temperatures onto Si (100) substrates (see Experimental section) exhibit a polycrystalline structure with the averaged grain diameter of 11.5±1.0 nm in the as-grown samples and of 15.0±1.0 nm after vacuum thermal annealing (VTA) for two hours at 673 K. Figure 2a shows a dark field micrograph with coherent scattering regions which can be associated with separate crystalline grains. The radial autocorrelation function (rACF) [52] of this micrograph is shown in the inset. To estimate the average grain diameter, the the rACF amplitude has been plotted as a function of rACF radius [52]. In Figure 2b, this dependence is shown for the as-grown state and after VTA.

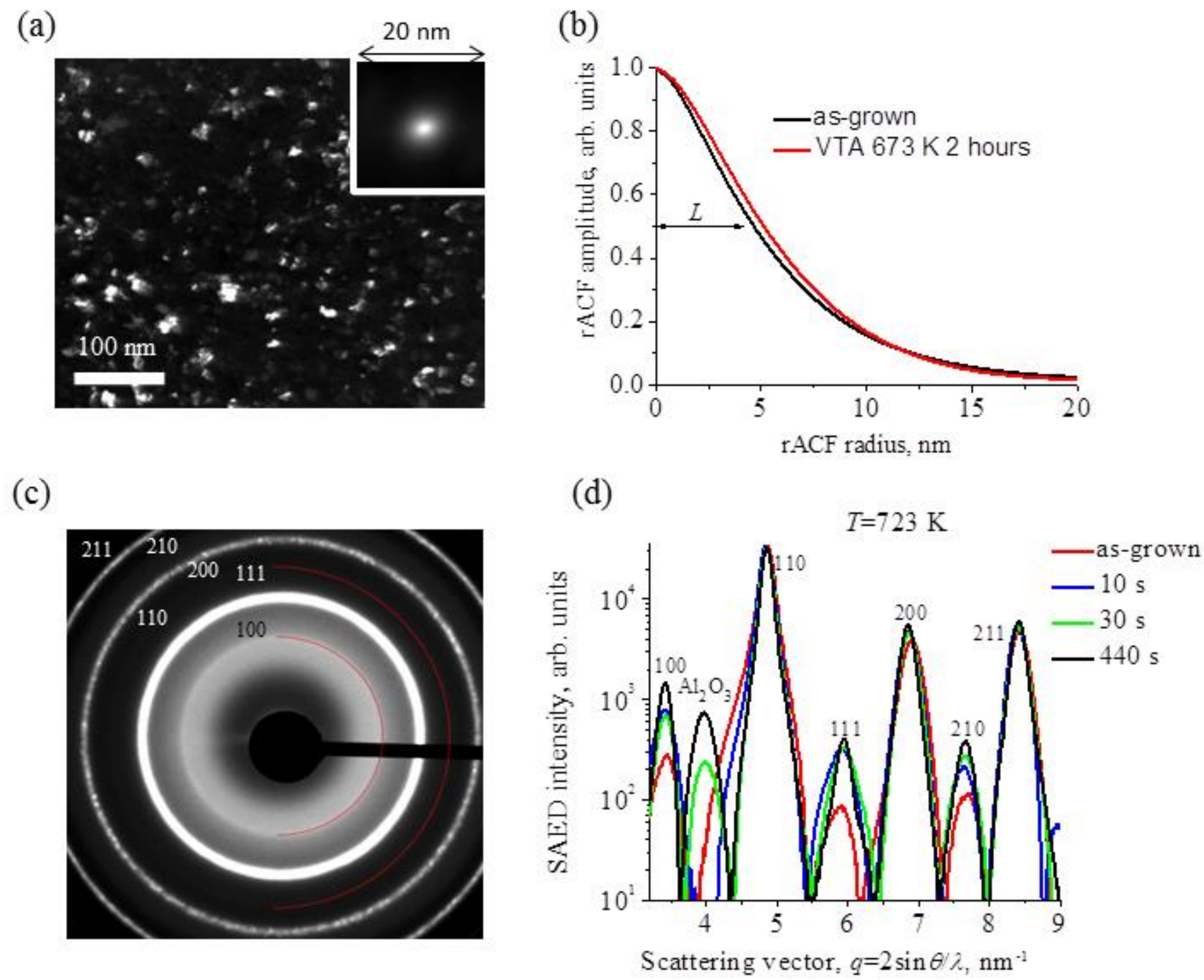

**Figure 2:** (a**)** A dark field TEM micrograph of an as-grown 50 nm thick Fe-Al sample onto a Si (100) substrate grown by magnetron sputtering at room temperature. The inset shows the radial autocorrelation function (rACF) [52] computed from the TEM micrograh. (b) The rACF amplitude versus the rACF radius, which gives the evaluation of averaged grain sizes. These can be estimated as 2.5×*L* [53], where *L* is full width at half maximum of the ACF rotational average. (c) Typical SAED patterns of Fe-Al subjected to VTA/RTA treatments. The superstructure diffraction peaks are highlighted in red. (d) Intensities of superstructure and fundamental diffraction peaks from the sample after subsequent RTA at 723 K. Note also the growth of the Al oxide phase during the RTA aging.

Using selected area electron diffraction (SAED) taken in the TEM mode, the enhancement of superstructure diffraction peaks – (100), (111), and (210) – has been observed after both VTA and rapid thermal annealing (RTA) treatments; see Experimental section. Obtained SAED data allowed us to estimate the chemical long-range order as [54]

$$S = \sqrt{(I_S/I_f)_{\exp}} \Big/ \sqrt{(I_S/I_f)_{theor}} \,, \qquad (1)$$

where $I_s$ and $I_f$ are experimental/theoretical intensities of the superstructure and fundamental diffraction peaks, respectively. Figures 2c and 2d show SAED patterns (c) and intensities of superstructure and fundamental diffraction peaks (d) after subsequent RTA treatments at 723 K. The theoretical intensities for superstructure diffraction peaks were taken for the completely ordered B2 structure in Fe-Al, where $(I_s/I_{110})_{theor}$=0.191, 0.034, and 0.03 for 100, 111, 210 superstructure diffraction peaks in $Fe_{50}Al_{50}$, respectively [55]. We find that in the as-grown samples there is partial ordering, $S(0)$~0.23. Note also the growth of the Al oxide phase during the RTA aging.

### 2.2. Laser Ablation Edge

Typically, nanosecond laser ablation of a thin metal film on a silicon substrate occurs at incident fluences above $I$~0.6 J/cm$^2$ [8]. A photograph in Figure 3a shows the ablation zone in a 50 nm thick Fe-Al sample (LQ sample) prepared on a Si substrate covered by a native $SiO_2$ layer. The pattern is obtained with a nanosecond laser pulse at air atmosphere. The pulse impinged on the sample at an energy of $E$=0.8 J along the normal direction to the sample plane. Such an irradiation is basically able to ablate the samples under study within the area in diameter of 0.80±0.05 cm. For a laser source employed in these studies [56], the spatial distribution of laser intensity over the sample surface can be approximated as

$$I(\rho) = I_0 \exp(-2\rho^2 / w^2), \qquad (2)$$

where $w$=0.5 cm is the beam radius of the used laser source. At the very center ($\rho$=0) of the ablation area, the laser intensity was $I_0 = 2E/\pi w^2$ =2.0 J/cm$^2$, while the ablation threshold occurring at $\rho$=0.4 cm was $I_{thr}$≈0.56 J/cm$^2$.

Figure 3b shows an optical micrograph of an LQ sample at the ablation edge. The sample was subjected to VTA (673 K for 2 h) before its LQ treatment (VTA+LQ sample). In this image one sees a dark line noted as A-A′, which separates two regions noted as “In” and “Out”. The region “In” is a periphery of the ablation zone, while the region “Out” is outside it. Figures 3c and 3d present typical secondary ion mass spectroscopy (SIMS) concentration profiles of Fe, Al, Si, AlO and O obtained in both regions. As seen from Figure 3b, SIMS probing was done in different locations (dark squares) around the A-A′ separation line. Only slight differences in the concentration profiles were found between locations 1, 2, 3, and 4 in the region “Out” and between locations 5, 6, and 7 in the region “In”. We see that remarkable changes occur in the concentration profiles at crossing the A-A′ separation line, namely diffusion intermixing of Si in the substrate with the film material (Fe, Al) inside the ablation zone (In). It is reasonable to suppose that the intermixing occurs in the melt. Note also that the melting temperature of Si (1687 K) is only slightly exceeds that of Fe-Al. It is also interesting that the onset of diffusion intermixing coincides with that of film evaporation. The latter fact has been revealed by taking the topographical relief at the ablation edge using optical profilometry. Figure 3e shows a two-dimensional (2D) topographical relief of this VTA+LQ sample. The topographical image taken is a counterpart of the optical micrograph presented in Figure 3b, with showing SIMS probed locations 3, 4, 5, and 6 only. Figure 3f depicts cross sections of the 2D image in Figure 3e along the dashed red and black lines. The arrows indicate the maximum slope in the topographical relief, which signifies the onset of laser-induced film evaporation. We find that the steepest topographical change occurs exactly at the A-A′ line. Therefore, it can be referred as the ablation edge.

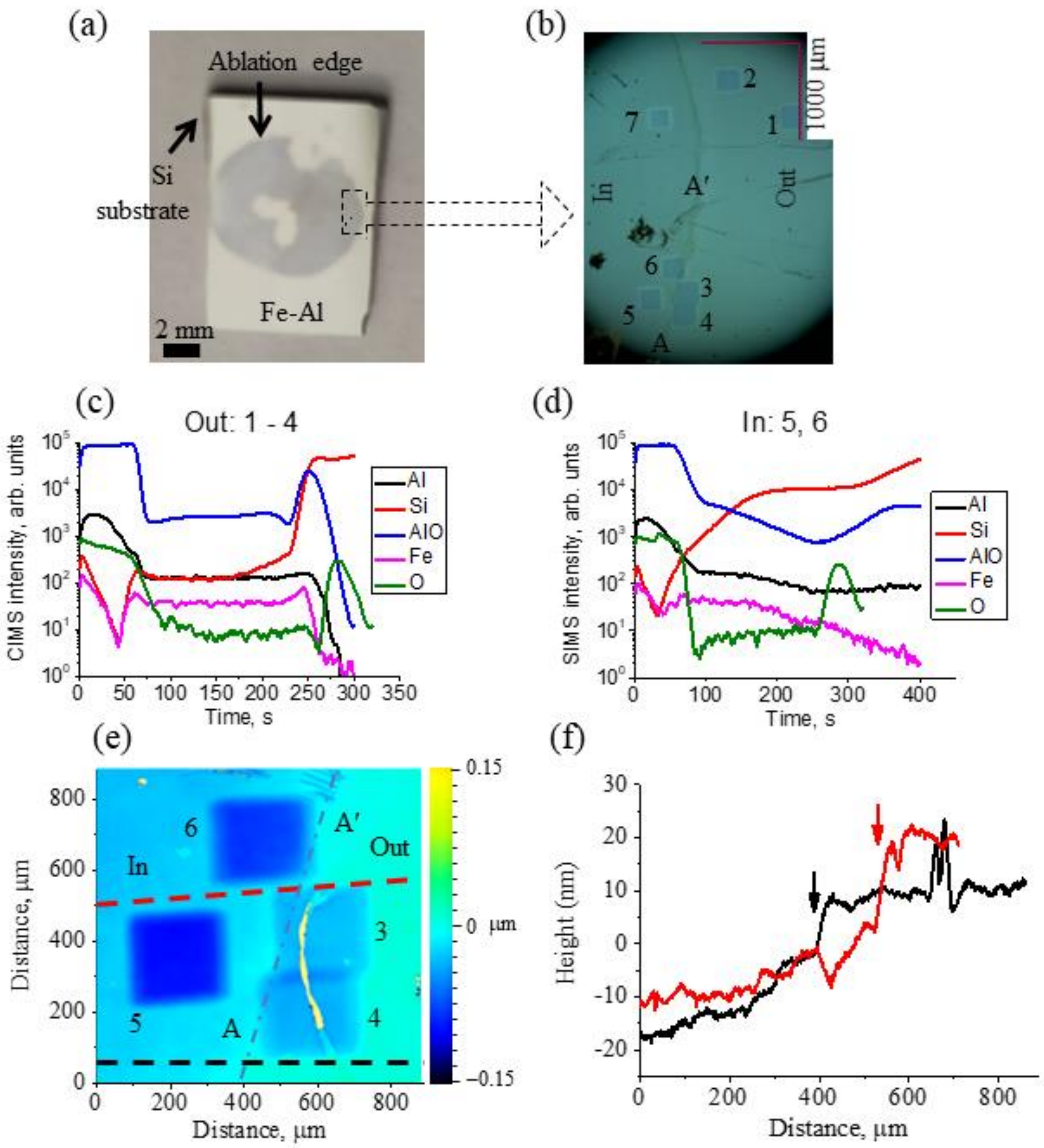

**Figure 3:** (a) Photograph of a 50 nm thick Fe-Al sample prepared onto a Si substrate with a native $SiO_2$ layer and irradiated by a nanosecond (5 ns) pulse from a Nd:YAG laser ($\lambda$=1064 nm) with intensity of $E$=0.8 J (LQ sample). (b) Optical micrograph of the ablation edge. The A-A′ line is associated with the onset of diffusion intermixing of the film material with Si substrate and with the onset of film evaporation. The dark squares noted as 1, 2, 3, 4, 5, 6, and 7 are locations of taken SIMS probes. (c, d) SIMS profiling in the LQ sample in different locations close to the A-A′ separation line, both outside the ablation zone (Out: 1-4) and inside it (In: 5-7). (e, f) Optical profilometry at the ablation edge: (e) Two-dimensional (2D) topographical relief which is a counterpart of the optical image (b), (f) Cross sections along the red and black dashed lines in the 2D image. The arrows indicate the maximum slope in the topographical relief, which signifies the onset of film evaporation.

Taking SIMS profiling (Figures 3c and 3d), we find the presence of ~10 nm thick surface AlO layer in all probed locations, even in the region "Out" far (location 1) from the ablation edge (A-A′ line) (Figure 3b). Formation of the aluminum oxide phase in our samples is supported by the selected area electron diffraction (SAED) data (Figure 2d). It is likely that the formed AlO covering plays a role of passivation layer, which prevents a sample subjected to high-temperature laser annealing from laser-induced oxidation. In Figures 3c and 3d, we also show the oxygen distribution across the film in a film region undisturbed by laser (location 1 in Figure 3b) and at the ablation edge (location 6 in Figures 3b and 3e).

## 2.3. Laser-induced Ferromagnetism

Figure 4a depicts a photograph of an LQ sample mounted into the MOKE setup and probed by MOKE. Scanning by the focused MOKE beam over the surface reveals that the maximum of MOKE response occurs at the very ablation edge (Figure 3a). As the A-A′ line (Figure 3b) heralds the start of Si melting and diffusion intermixing between the film and substrate (Figures 3c and 3d), it is likely that the strongest MOKE is exactly at this separation line, i.e., ablation edge (Figure 3b). It is interesting that the MOKE maximum in LQ samples is significantly higher than that in as-grown ones. Figure 4b shows a magnetic hysteresis loop in this maximum. For comparison, magnetization curves of the as-grown sample and after its vacuum thermal annealing (VTA) for two hours at 673 K are presented in the same plot. The enhancement of the MOKE response $2\theta_s$ induced by laser is associated with full disordering in the transient melted state. The disorder ($S\rightarrow 0$) persists in the solidified alloy [8] after terminating the laser pulse. It was also remarkably to find an enhancement of the magnetic coercivity up to the factor of ~20 in the region of enhanced MOKE response. To our knowledge, magnetic hardening induced by short-pulse laser irradiation was not reported hitherto.

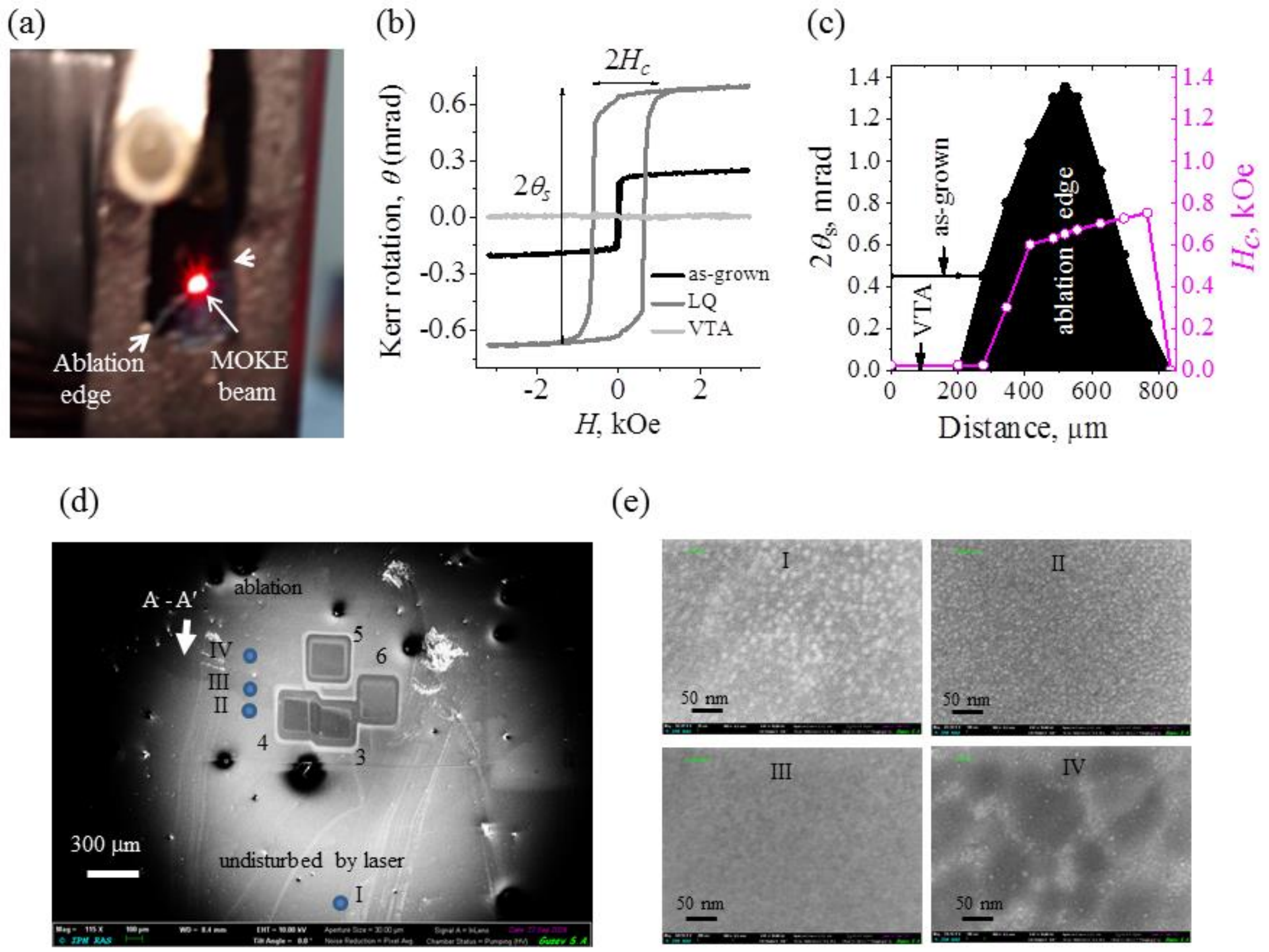

**Figure 4:** (a) A photograph of an LQ sample mounted into the MOKE setup and probed with MOKE. The MOKE response is found to enhance exactly at the laser ablation edge, which is identifiable optically as the start of diffusion intermixing between the substrate and film (Figure 3d). (b) The Kerr rotation angle $\theta$ versus the magnetic field $H$ applied in the sample plane. The magnetic hysteresis loop is taken in the maximum of the MOKE response 2$\theta_s$ occurring at the ablation edge. For comparison, magnetization curves are shown for the as-grown sample and after vacuum thermal annealing (VTA) for two hours at 673 K. The VTA caused reducing 2$\theta_s$ to a value that is beyond the accuracy limit of the MOKE measurements. (c) Spatial distribution of 2$\theta_s$ and coercivity $H_c$ along the radial direction. (d) SEM image of the VTA+LQ sample with indicating the SIMS probed locations (dark squares) 3, 4, 5, and 6. (e) High-resolution SEM images taken in locations I, II, III, and IV pointed out in the micrograph (d).

Another feature we reveal in our samples, namely reducing the magnetization under thermal annealing at moderate $T$, was earlier reported [8, 9, 42] to be explained in terms of thermally induced chemical ordering and formation of the B2 state. In our work, such an interpretation of the MOKE behavior and magnetization is supported by the SAED data shown in Figures 2c and 2d.

As for the spatial distributions of $2\theta_s$ and $H_c$, Figure 4c shows these quantities as functions of a distance $\rho$ along the radial direction. It is reasonable to assume that the LQ treatment used produces a ring-shaped LQ region around the ablation zone. [8] As seen also from Figure 4c, the magneto-optical/magnetic properties of the solidified alloy within this ring differ from those of the as-grown sample undisturbed by the laser. It is also likely that these are affected by diffusion intermixing in the melt between the Fe-Al film and Si substrate, which is observed with SIMS profiling (Figures 3c and 3d).

It here can be important to note that, at the very ablation edge, the polycrystalline structure characteristic for the as-grown samples (Figures 2a and 2b) vanishes and the film becomes nearly amorphous. Figures 4d and 4e show scanning electron microscopy (SEM) images at the ablation edge of the same sample whose optical and topographical images were shown in Figures 3b and 3e, respectively. If the image in Figure 4d is a general view of the ablation edge, Figure 4e presents high-resolution SEM images taken in different film locations: (I) far outside the A-A′ transition zone (undisturbed by laser), (II) still outside but close to the ablation edge, (III) inside the A-A′ transition zone, and (IV) inside the ablation zone. We see that the granular structure vanishes in the location III at all. However, the film microstructure becomes heterogeneous again, with much a bigger scale of heterogeneities, inside the ablation zone (IV), which is likely to be due to laser dewetting [57].

### 2.4. Ordering Kinetics

In these studies, the goal is to compare the thermally induced $S(t)$ ordering kinetics for as-grown and LQ samples. This comparison can provide information about the behavior of vacancies and their supersaturation, as the ordering is believed to occur through point defects such as vacancies in the atomic lattice. The vacancy supersaturation can be estimated by fitting a theoretical model for $S(t)$ evolution to that dependence obtained experimentally. However, straightforward $S(t)$ measurements in LQ samples with micron-scale modified regions (Figure 4c) are beyond of the work we report here. We have investigated magneto-structural correlations [45, 46] between $S$ and $2\theta_s$ in the as-grown samples which were subject to rapid thermal annealing (RTA) without LQ treatment. In these studies, our assumption is that still the same relationship between $S$ and $2\theta_s$ would retain in LQ samples, in which $S(0)\rightarrow 0$ after melting and solidification at the ablation edge. Importantly, this hypothesis is verifiable by extrapolation into the $S\rightarrow 0$ region of the $S$-versus-$2\theta_s$ dependence retrieved in the as-grown samples. Remarkably, the as-grown samples are found to be partially chemically ordered (S=0.23) even at $t$=0 (Figure 2d).

Figure 5a shows $\theta_s(t)/\theta_s(0)$ and $1–S(t)+S(0)$ in the as-grown sample under RTA aging at 723 K. A good correlation is observed between temporal changes in $\theta_s$ and $S$ during the relaxation process. For example, the RTA for a time of $t$~10 s produces noticeable changes in both $2\theta_s$ and $S$. Increasing the RTA aging duration to $t$~30 s at the same $T$ does not cause changes in either $2\theta_s$ or $S$. Therefore, the while needed for heating the sample up to set $T$=723 K is rather not longer than ~10 s. We also find that a subsequent increase in RTA aging duration to $t$~500 s provides a further decrease in $2\theta_s$ and $1–S+S(0)$. The data presented in Figure 5a allow us to establish the relationship between $S$ and $2\theta_s$, which is shown in Figure 5b taking into account that $S(0)\rightarrow 0$ in LQ samples.

In the following, LQ samples were aged using RTA at different annealing temperatures $T$, while monitoring the ordering kinetics using MOKE. It is interesting that a preliminary VTA aging for 2 h at 673 K followed by LQ treatment provided a relatively fast ordering kinetics under subsequent RTA aging of such VTA+LQ samples. Figures 5c and 5d show representative examples of the observed time evolution of the MOKE response $\theta_s/\theta_s(0)$ and magnetic coercivity $H_c/H_0$ ($H_0$~0.03 kOe is the coercivity of the as-grown sample) in VTA+LQ samples subjected to RTA at 583 K and 423 K. For comparison, the time dependences of the MOKE response obtained for the as-grown samples are also presented. The numbers on the plots correspond to values of $S$ obtained from Figure 5b. We see that RTA aging of VTA+LQ samples results in a comparatively stronger decrease in $2\theta_s$ and, consequently, in a stronger increase in $S$. For example, after RTA aging during $t$~$10^3$ s at 583 K, the MOKE maximum in the VTA+LQ samples becomes even lower than the MOKE intensity in the as-grown sample aged under the same conditions. Nevertheless, the initial ($t$=0) MOKE response in LQ samples (either with preliminary VTA or without it) is greater than that in the as-grown samples due to laser-induced chemical disordering. In VTA+LQ samples, the change in $2\theta_s$ observed corresponds to $S–S(0)$~0.47. By comparison, the as-grown sample exhibits a weaker ordering under RTA aging, $S–S(0)$~0.2.

It is also interesting that, in LQ samples, the temporal behavior of the coercivity $H_c$ under RTA aging depends on annealing temperature $T$. At 583 K, for example, $H_c$ initially strongly decreases during a while of $t$~10 s, as seen from the inset of Figure 5c. However, this magnitude even increases under a longer RTA aging at this $T$. Unlike the behavior of $H_c$ at 583 K, we do not find a decrease in $H_c$ under RTA aging at a lower $T$=423 K, which is illustrated in the inset of Figure 5d.

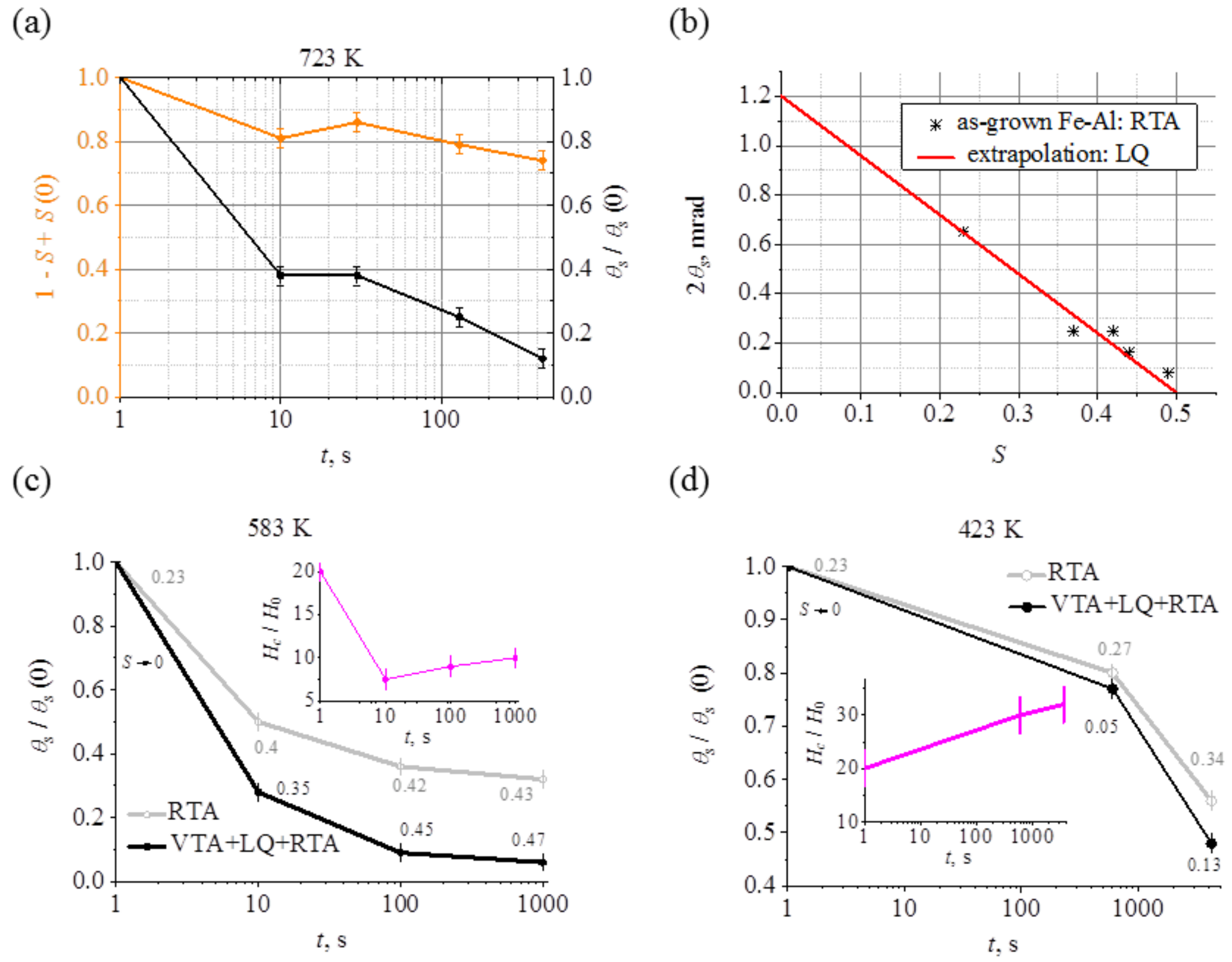

**Figure 5:** (a) The long-range order $S$ and MOKE response $2\theta_s$ versus time $t$ under RTA at 723 K. (b) Relationship between $2\theta_s$ and $S$, which has been found from the ordering kinetics illustrated in plot (a). The red line is extrapolation for an LQ sample, with taking into account that $S \to 0$. This extrapolation is compatible with $2\theta_s$ experimentally found in LQ samples. (c-d) $2\theta_s$ versus $t$ under aging at 583 K and 423 K in the as-grown (in blue) and LQ (in orange) samples. The LQ sample was subjected to VTA for 2 hours at 673 K before LQ treatment. The numbers on the plots are values of $S$ for as-grown and VTA+LQ samples at $t$=0 s, 10 s, 100 s, and 1000 s, which have been found from the relationship between $2\theta_s$ and $S$ in as-grown samples. The insets show the temporal evolution of the magnetic coercivity $H_c$ measured after LQ treatment, which was normalized to that quantity ($H_0$) in the as-grown sample (without VTA treatment).

### 2.5. Evolution of Vacancy Supersaturation

To estimate the concentration of excess vacancies, the ordering kinetics observed experimentally was compared to that derived by solving numerically the chemical balance equation proposed by Dienes [51], taking into account the time dependence of vacancy concentration [40]

$$c_V(t) = c_{eq} + (c_V(0) - c_{eq})\exp(-t/\tau) \tag{3}$$

with $\tau$ being the characteristic time for vacancy annihilation at their sinks:

$$\frac{dS}{dt} = \frac{D_{S=0}(T)}{a^2}\left[1+(r-1)\exp(-t/\tau)\right]\exp\left[-\frac{T_c S}{\gamma T}\right]\times$$
$$\times\left\{\gamma(1-S)^2(\exp\left[\frac{T_c S}{\gamma T}\right]-1)-S\right\}, \qquad (4)$$

where $D_{S=0}(T)=D_m(T)c_{eq}(T)$ is the interdiffusion coefficient in a disordered alloy ($S$=0) at $c_v \to c_{eq}$, $c_{eq}(T)=\exp(s_v/k)\exp(-E_v/kT)$, $s_v$ and $E_v$ are the vacancy formation entropy and energy, respectively, $D_m=D_0\exp(-U/kT)$ is the vacancy mobility at $S$=0, $D_0=\nu a^2\exp(s_m/k)$, $\nu$ is the vibration frequency of atoms in the lattice, $s_m$ is the activation entropy of atomic jump from a lattice site occupied by the atom to a neighboring one, $a$ is the lattice constant, $\gamma=x(1-x)$ with $x$ being the composition of a binary alloy $A_xB_{1-x}$, and $r=c_v(0)/c_{eq}(T)$ is the vacancy supersaturation rate at $t$=0.

According to Equation (4), a simulated $S(t)$ evolution has to be fitted to the experiment on the four material parameters, namely $D_{S=0}$, $\tau$, $S(0)$, and $r$. Figure 6a shows the theoretical dependences obtained by fitting to the experimental dependences for as-grown and VTA+LQ samples aged at $T$=583 K (Figure 5c). Both theoretical dependences plotted in Figure 6a have been simulated at $\ln D_{S=0}= -11.5$ (nm$^2$/s) and $\tau$=5.0 s. As for the initial value ($t$=0) of the long-range order $S$, this parameter taken from the experiment is $S(0)$=0.23 in the as-grown samples (Figure 2d) . However, it becomes close to zero in LQ samples due to melting of the film at the ablation edge. It is found that the simulated evolution is well fitted to the experiment at $r\sim 3.3\times 10^3$ and $r\sim 1.4\times 10^4$ for as-grown and VTA+LQ samples, respectively, and so, VTA+LQ samples may exhibit a vacancy supersaturation, which is about four times larger than that in the as-grown samples. The evolution of vacancy concentration $c_v$ in the as-grown and VTA+LQ samples, which is calculated in accordance with Equation (3) at $\tau$=5.0 s ($T$=583 K), is shown in Figure 6b. For this plotting, a value of $c_{eq}(T=583\text{ K})= 5.0\times 10^{-7}$ is taken from Ref. [41], where $c_{eq}(T)$ measured in Fe-Al were reported at different $T$. Thus, the absolute value of $c_v(0)$ estimated in VTA+LQ samples is as high as ~0.01, which can be compared to $c_{eq}$ in the system at high $T\sim T_m$. It here should be noted that, at $S(0)\to 0$ in LQ samples, the ordering strongly depends on how exactly $S(0)$ is close to zero. This feature is illustrated in Figure 6c, in which the temporal evolution of $S(t)$ is computed at different $S(0)$ – $10^{-2}$, $10^{-3}$, and $10^{-4}$. It is found that correct fitting to the experiment can be performed at a single combination of $D_{S=0}$, $\tau$, and $S(0)=10^{-4}$.

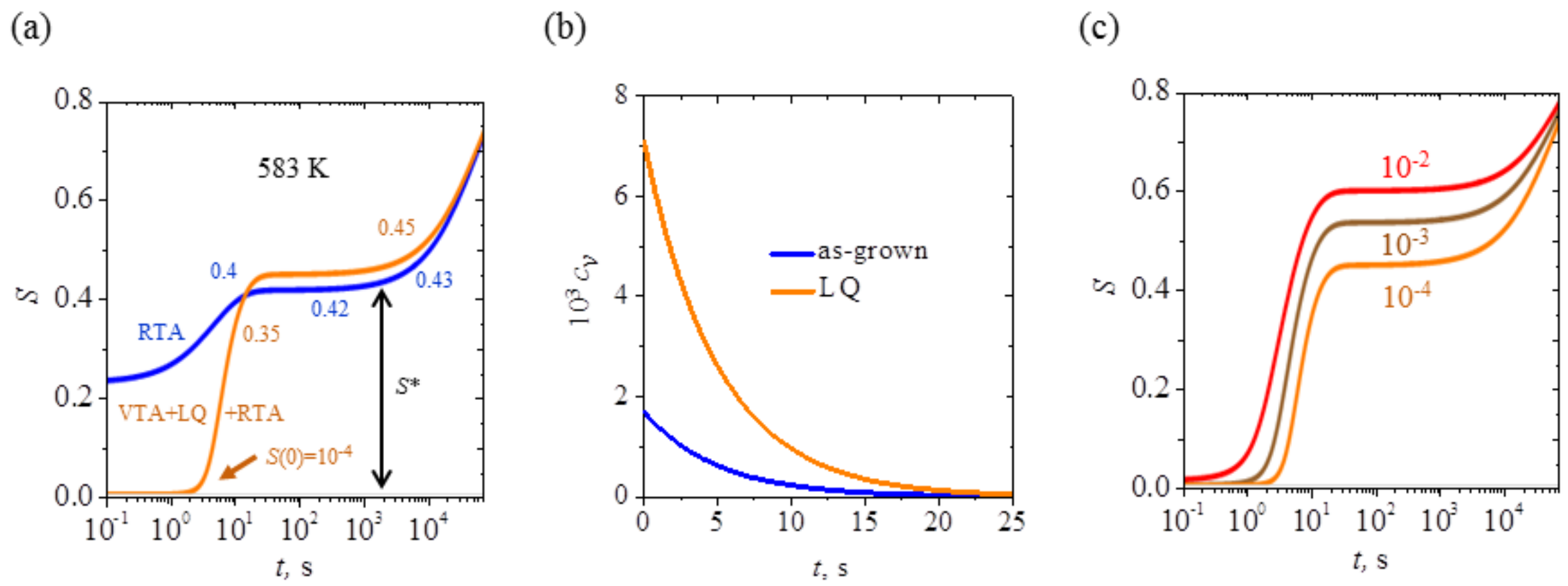

**Figure 6:** (a) Temporal evolution of $S$ simulated at $T$=583 K. The blue and orange curves have been plotted by fitting to the ordering kinetics obtained experimentally (Figure 5c) for as-grown and VTA+LQ samples, respectively. The values of $S$ indicated in the plot are computed at $t$=10 s, 100 s, and 1000 s. These values are equal to those obtained experimentally. (b) Vacancy concentration $c_v$ versus $t$ in the as-grown and LQ samples. (c) Illustration of how strongly the evolution of $S$ depends on $S(0)$. The numbers $10^{-2}$, $10^{-3}$, and $10^{-4}$ are values of $S(0)$ close to zero, at which the data were simulated.

The fitting of the theoretical evolution to the experimental one, illustrated in Figure 6a, was done by taking into account the following considerations. First, note that the temporal evolution of $S(t)$ is two-step [40]. The first step occurring for a shorter timescale $t$~10 s is associated with excess vacancies in the atomic lattice [16], while the second one ($t$~$10^5$ s) results from reactivation of the ordering process with the characteristic timescale which is defined by only $D_{S=0}$ when $c_v(t)\rightarrow c_{eq}(T)$ [16, 40]. The short-time ordering is characterized by the time $\tau$ necessary for reaching the asymptotic value $S$* when excess vacancies annihilate at their sinks. The step height $S$* depends on other parameters, namely $D_{S=0}$, $S(0)$, and $r$.

Experimentally, we are able to follow the short-time ordering only under RTA aging for times up to ~$10^4$ s. We find that ordering trends at 583 K and 423 K are quite different, as shown in Figures 5c and 5d for both the RTA and VTA+LQ+RTA samples. These differences are associated with the strongest parameter affecting the rate of ordering, namely $D_{S=0}(T)$ which is contained in Equation (4) and obeyed to the Arrhenius law. It is also likely that the decrease in the magnitude of $H_c$ under RTA aging at $T$=583 K for $t$~10 s (inset of Figure 5c) results from migration and annihilation of excess vacancies in the atomic lattice. However, at a lower $T$=423 K, the vacancies are "frozen" in the lattice and thus cannot mediate a decrease in $H_c$ of LQ samples (inset of Figure 5d). The enhancement of vacancy concentration can give even plastic deformation of a magnetic material and thus can modify reversal magnetization modes therein [28].

## 3. Conclusion

A thin (50 nm) film of near equiatomic Fe-Al alloy sputtered onto a Si (100) substrate is found to exhibit acceleration in chemical ordering after irradiation of the alloy with nanosecond laser pulse (LQ treatment). The data were obtained with magneto-optical Kerr effect (MOKE) magnetometry by finding preliminarily the relationship between the MOKE response and the long-range atomic order $S$ in the as-grown samples subjected to aging using rapid thermal annealing (RTA). Temporal ($t$) dependences of $S$ retrieved experimentally were compared to those found by numerical solution of the Dienes equation [51] for ordering kinetics, taking into account the time-dependent vacancy concentration, $c_v(t)$ [40]. For LQ samples, the theoretical evolution can be fitted to the experiment by assuming that the vacancy concentration is about four times higher than that in as-grown samples. The LQ effects have been observed at the very ablation edge, which is well indicated optically and with scanning electron microscopy. It is likely that vacancies in the atomic lattice produced with a sufficient density mediate the observed magnetic hardening in LQ samples. Thus, the results reported here will open a room for studying excess vacancy behavior in crystals within a small (micron-scale) volume and their effects on physical properties, e.g., magnetic coercivity in Fe-Al.

## 4. Experimental Section

***Sample preparation***: Near equiatomic Fe-Al alloys were prepared as thin films with a thickness of 50 nm using dc magnetron co-sputtering from individual Fe and Al targets onto Si (100) substrates with a native $SiO_2$ surface layer. In these experiments, an AJA 2200 multichamber system was used at a basic pressure down to ~$6\times10^{-8}$ mbar. The film composition was checked by measuring the critical angle for X-ray total reflection using a Bruker D8 Discover x-ray diffractometer. To study the phase transformations and their kinetics, the prepared samples were subjected to nanosecond laser annealing [laser quenching (LQ)] at air atmosphere. Besides, the samples (either with LQ treatment or without it) were annealed isothermally either in a furnace installed in a vacuum chamber (VTA) or by using a rapid thermal annealing (RTA). VTA treatments were performed at a residual gas pressure of ~$4\times10^{-6}$ mbar. The RTA system operated with He gas flowing through a heated sample at a flow rate of 10 l/min and a pressure of 1.5 kg/cm$^2$. Unlike VTA, RTA treatments allow for short-term (down to $t$~10 s) thermal annealing. Experiments on LQ treatments were carried out with a nanosecond (~5 ns) Q-switch Nd:YAG laser (EKSPLA NL300 series) operating at the fundamental wavelength of radiation ($\lambda$=1064 nm). [56]

***Sample characterization*:** The process of chemical ordering occurring in the films under RTA/VTA treatments was studied with selected area electron diffraction (SAED) in the transmission electron microscopy (TEM) mode and magneto-optical Kerr effect (MOKE) magnetometry. Taking SAED data from the samples prepared on commercial 50 nm thick $Si_3N_4$ membranes allowed for determining the long-range order $S$. The TEM/SAED studies were carried out using a transmission electron microscope LIBRA 200 MC (Carl Zeiss, Jena) operated at 200 kV. Besides, the surface morphology of the samples under study was examined with a Carl Zeiss Supra 50 VP scanning electron microscope (SEM) and optical profilometer Talysurf CCI 2000. The formation and properties of micron-scale zones under LQ treatment were investigated by MOKE magnetometry in combination with secondary ion mass spectroscopy (SIMS) at a TOF.SIMS-5 time-of-flight mass spectrometer. The used MOKE setup was home-built and based on a Faraday modulator technique. As a light source, a He-Ne laser ($\lambda$=633 nm, 5 mW, Thorlabs HRS015B) was employed. The MOKE intensity was measured at *room temperature* as a function of magnetic field $H$ applied in the film plane to generate a MOKE hysteresis loop. The accuracy of measuring the Kerr rotation angle $\theta$ in this MOKE setup was as high as ~$0.2\times10^{-4}$ rad. To detect the CODTs inside the narrow ring formed by LQ treatment, the MOKE beam was focused onto the ~50-μm spot in the sample surface. For SIMS profiling, the sputtering was performed by $Cs^+$ beam (50-µm width) with ion energy of 1 keV at an incidence angle of 45°; the sputtering raster was 200 ×200 μm (indicated in the optical images) The probing was performed by $Bi^+$ beam (2-µm width) with ion energy of 25 keV. The probing raster was chosen to be 100 × 100 μm.

## Acknowledgements

**W**ork was supported by the Russian Science Foundation (grant number 23-22-00044). In the studies, the facilities of the Common Research Centrum «Physics and technology of micro- and nanostructures» (IPM RAS) was employed. The authors thank Evgeny V. Skorokhodov and Pavel A. Yunin for taking the SEM and optical profilometry data, respectively.